# RF/MW power standard realization without unitary efficiency assumption at dc/LF

Luca Oberto[1], Luciano Brunetti[1]

[1]Istituto Nazionale di Ricerca Metrologica – INRIM
Strada delle Cacce, 91, 10135 Torino, Italy
E-mail: l.oberto@inrim.it

*Abstract* — **Thermoelectric power sensors are widely used in commercial power meters at RF and microwaves, due to their superior robustness, stability, and accuracy if compared with other types of power sensors. Furthermore, their electrical architecture and related performance turned out to be very useful in the realization of the broadband primary power standards as alternative to resistive power sensors, i.e. bolometers. Hereby we present a comparison in term of effective efficiency of a thermoelectric sensor calibrated by applying two different methods of power substitution when used as thermal load in a coaxial microcalorimeter at RF and microwaves. The aim is to test a technology that could enable the realization of a primary RF/MW power standard independently of the assumption of unitary efficiency at the dc (or LF) reference power.**

## I. INTRODUCTION

Realization of primary power standards at radio frequency and microwave (RF and MW) has always been an important task for National Metrology Institutes (NMIs), and also a real challenge. The method used by most NMIs consists in calibrating a thermal detector against a *dc power standard* by means of the power substitution method applied in calorimetric measurements [1], [2]. The process dates back to the late 1950's and still today it is basically the same, even though several technical improvements have been introduced during the last decades [3] – [12]. A significant change consists in the use of the thermoelectric detection rather than original bolometric detection. This has been suggested due to the lack of commercial bolometric sensors, but is also motivated by the superior performance of thermoelectric sensors, as it has been demonstrated for the coaxial case at least, [4], [12], [13]. Recently a noticeable W-band waveguide power standard has been realized by using a thermoelectric sensor that is fed via a waveguide to coplanar waveguide transition line [14].

However, from our point of view, the most interesting characteristic of the device used in [14] consists of an additional auxiliary heater, placed near the RF/MW load, but electrically insulated from it. Sensors exploiting this feature have been recently introduced with coaxial input line also [15].

The original purpose of such technical solution is to provide a possibility for sensor power meter self-calibration. De facto, it creates an alternative way for supplying the reference power into microcalorimeters that use thermoelectric power sensors as thermal loads. Currently, such microcalorimeters are fed through the RF/MW input coaxial line only [4]. In few words, the *effective efficiency* of the device mentioned in [15] can be determined at each frequency in two different ways provided that the auxiliary heater is in good thermal contact with the RF-MW absorber.

This paper evaluates how much the effective efficiency values obtained by two different power substitution methods are consistent to each other. If the auxiliary heater will turn out to be usable, this will enable a realization of a primary RF/MW power standard that no longer requires the assumption of unitary efficiency at the dc or low frequency (LF) reference power and its experimental verification [16]. Conversely, by feeding the sensor with the reference power through the coaxial input, one have to assume that no losses exist at the reference power frequency (that is, unitary efficiency at dc or LF) [16].

The paper presents the calibration of a power sensor of the type mentioned in [15] at some selected frequencies in the 0.01-18 GHz range. Measurements have been performed with the INRIM coaxial microcalorimeter by using two different power substitution schemes as described and commented in the following sections.

To test the reproducibility of this result, more measurements on different sensors have been planned. Anyway we are confident on the outcome of our comparison since the production technology shows already enough reliability to be used in sensors launched on the market [15].

## II. DEVICES AND THEORY

In our experiment we consider a new type of commercial thermoelectric sensor having precision connector type-N (Navy, i.e. PCN connector), which has been provided with an auxiliary heater directly by the manufacturer [15]. The sensor architecture allows us to perform two different calibrations when we use the device as microcalorimeter load for realizing a primary power standard at RF and MW. Figures 1 and 2 show the sensor mount realized by INRIM to make the original device compatible with the 7 mm coaxial microcalorimeter.

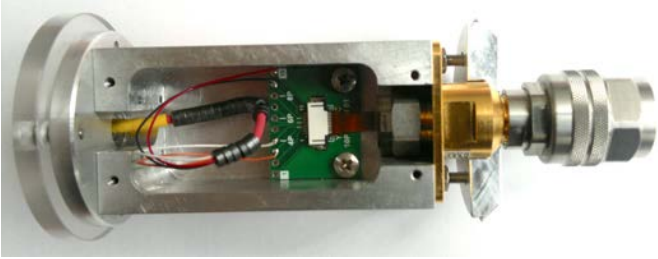

Fig. 1. Picture of the specific sensor mount under test.

The quantity we want to measure is the effective efficiency $\eta_e$ of the sensor mount, which can be defined, in the most generic way, as the ratio of measured power $P_M$ to total absorbed power $P_A = (P_M + P_X)$:

$$\eta_e = \frac{P_M}{P_M + P_X}, \qquad (1)$$

where $P_X$ represents the total power loss in the sensor mount [4]. For the thermoelectric sensors used for our measurements, $\eta_e$ has been introduced in a different manner [3], that is:

$$\eta_e = \left.\frac{P_{dc}}{P_{RF}}\right|_{V_{dc}=V_{RF}=\text{constant}}. \qquad (2)$$

Definition (2) is given as the ratio of the reference dc-power $P_{dc}$ to total RF-MW power dissipated in the mount $P_{RF}$, each one producing the same sensor thermopile response $V_{dc} = V_{RF}$. Equation (2) implies that the following condition holds:

$$\frac{P_{dc}}{H_{dc}} = V_{dc} = V_{RF} = \frac{P_{RF}}{H_{RF}}, \qquad (3)$$

where $H_{dc}$ and $H_{RF}$ are quantities dependent both on thermal properties and on the electromagnetic field distribution inside the feeding line and power sensor assembly [7], [9], [13]. The effective efficiency $\eta_e$ can be expressed in term of previous quantities:

$$\eta_e = \frac{H_{dc}}{H_{RF}}. \qquad (4)$$

Evidently $H_{dc}$ in not equal to $H_{RF}$ because of different distribution of the electromagnetic field in the feeding line and absorber block, but while $H_{RF}$ is unique, $H_{dc}$ could differ if power substitution is made via the RF-absorber or via the auxiliary heater.

If $H_{dc}$ is reasonably not sensitive to the change of the power substitution paths, then the main absorber and auxiliary heater are both in good thermal contact with the sensor thermometer. If that stands, Eq. (4) returns a $\eta_e$-value independent of the power substitution way selected.

Unfortunately, (4) is not much useful for the measurand determination, because both $H_{RF}$ and $H_{dc}$ are not easily related to the physical quantities we can measure using a microcalorimeter. Therefore we must proceed by using the technique previously developed at INRIM for realizing a RF/MW power standard based on thermoelectric sensor in coaxial line [4], [12], [13]. We demonstrated the congruence of (3) with (1) and, at the same time, from the generic definition (1) we deduced an equation that allows measurement of $\eta_e$, that is:

$$\eta_e = \left(\frac{e_2}{e_1}\right)\left[1 - \left(1 + |\Gamma_S|^2\right)\frac{e_{1SC}}{2e_1}\right]^{-1}, \qquad (5)$$

where $e_1$ and $e_2$ are the electrical responses of the microcalorimeter to the RF/MW power and to the reference power (dc or 1 kHz in our case) substituted into the system, respectively. The condition $V_{dc} = V_{RF}$ has also to be maintained as in (2). The voltage $e_{1SC}$ corrects for the microcalorimeter response due to system losses and is determined by means of the short circuit technique [4], [16]. We stress that the system losses are dominated by the RF/MW losses in the microcalorimeter insulating line section. Finally, the term $(1+|\Gamma_S|^2)$ is an additional correction necessary to enhance the result accuracy when the reflection coefficient $\Gamma_S$ of sensors under calibration is not negligible [17]. A further expression of $\eta_e$ exists that takes into account possible losses into the specific short circuit used to calibrate the microcalorimeter [16]. However in our case the short circuit losses have been verified to be negligible with measurements that returned a short circuit reflection coefficient close to the ideal value.

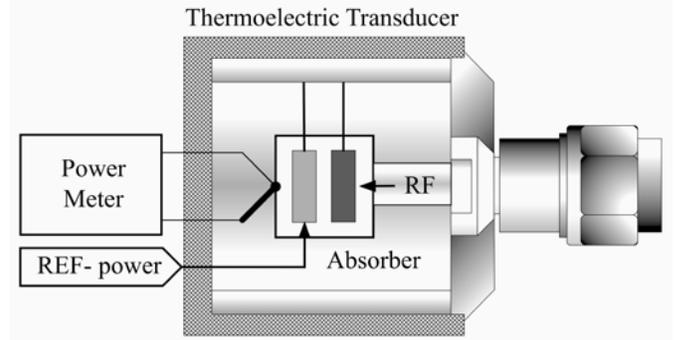

Fig. 2. Scheme of the power sensor considered for the alternative power standard realization described in this work.

A critical point of this measurement technique is that it requires the assumption of unitary efficiency of the sensor at dc or LF. This is necessary if the power substitution is made on the RF coaxial line, otherwise it is not possible to obtain an operative equation of effective efficiency [16], [17]. The assumption has to be verified by difficult measurements, as at the reference power (dc or LF), possible losses are close to the instrument sensitivity. The equivalence of the mentioned power substitution methods, if demonstrated, allows the realization of the microwave primary power standard independently of the cited assumption and its verification. This because the power substitution can be done on the auxiliary heater line, where possible losses at dc or LF do not influence the effective efficiency determination [12], [13].

## III. MEASUREMENTS AND RESULTS

Experimental work consists of the calibration of a specific thermoelectric sensor in 7 mm coaxial line at several frequencies in the band 0.01 – 18 GHz. For the purpose, the test port of INRIM coaxial microcalorimeter has been fitted with precision connector type-N (Navy, i.e. PCN connector). The power sensor under test has been obtained by modifying a commercial device by removing all its electronic components in order to have direct access to the sensitive element, and inserting it in a custom made case, so to obtain an appropriate microcalorimeter thermal load.

Calibrations have been performed according to the INRIM procedure, which requires to operate in a shielded and thermostatic environment at $(23 \pm 0.3)°C$ with relative humidity of $(45 \pm 5)\%$, in order to obtain the necessary system stability and repeatability. The same procedure suggests to apply a reference power at 1 kHz rather than in dc, to avoid effects of spurious thermoelectric voltages. This operation is allowed because it has been proved that the system losses at 1 kHz are of the same order of the dc losses, if any when supplying the reference power via the input coaxial line [16].

Firstly, the power sensor under calibration is supplied by a nominal power of 1 mW at 1 kHz for a time period of 400 min. Then, the reference power is substituted with an appropriate RF/MW power level so to maintain constant the sensor output signal for the next 400 min. Finally the RF/MW power is switched off and substituted again with the reference 1 kHz power. Afterward the RF/MW frequency can be set to a new value and another measurement cycle begins. The commutation time of 400 min between 1 kHz and RF/MW power, is given by the microcalorimeter time constant. It has been experimentally found to allow the system to reach the thermal equilibrium state after each power substitution.

The described measurement protocol is always applied independently of the power substitution method and also in the short circuit calibration measurements.

In any case, for each power substitution step, the microcalorimeter thermometer, i.e. a thermopile placed at the connector base of the sensor under test, generates a dc signal $e$ of the order of some dozens of μV, depending on the power level supplied to the system and on the system losses. When RF/MW is supplied to the calorimetric load, the system thermometer measures an exponential temperature increase on the same load up to an asymptotic value $e_1$. Conversely, when RF/MW is substituted by 1 mW reference power at 1 kHz, there is a cooling of the thermal load resulting in a decrease of the asymptotic value $e_2$.

Asymptote $e_1$ is proportional to the parasitic losses at RF/MW of calorimetric load, whereas the asymptote $e_2$ is a measure of the equilibrium temperature the system reaches in absence of losses. The typical form of this signal is shown in Fig. 3. In the same frame there is the trend of the sensor output that is a measure of how well the power supplied to the system is stabilized. This sensor response complies well with (2), because the substitution error of the reference power at 1 kHz with the RF/MW power turned out to be negligible, supporting the validity of (5).

A fitting process based on the Levenberg – Marquardt algorithm has been used to obtain the asymptote values $e_2$, $e_1$ and $e_{1SC}$ from measurements, together with the uncertainty component related to the thermometer accuracy [18] to be used in (5).

Table 1 shows two values of $\eta_e$ for each measurement frequency so to highlight what discrepancy exists between the two methods of power substitution. Their associated uncertainties arise from the above mentioned thermometer accuracy, the thermal instability of the system and the sensor reflection coefficient. The whole measurement uncertainties associated to the measurand $\eta_e$ have been calculated by applying the Gaussian error propagation on (5), to be conformal with [19]. It is worth noting that the short circuit measurements are made by short-circuiting the input connector of the DUT by means of a thin conductive foil of INRIM design. Its thermal mass is very small with respect to that of the DUT and of the coaxial feeding line, so it does not have any measurable influence on the thermal properties of the system. This method proven to be effective with dedicated measurements and is routinely used at INRIM since many years. No thermal imbalance has been observed between matched and short circuit measurements, within the instrumental sensitivity. Anyway, both matched and short circuit measurements are performed at every frequency and several measurement cycles may be repeated to check the reproducibility contribution, which turned out to be negligible.

In Table 1 the compatibility index is also shown for ease of comparison. It is defined as follows:

$$E_n = \frac{|\eta_{e1} - \eta_{e2}|}{\sqrt{U(\eta_{e1})^2 - U(\eta_{e2})^2}}. \qquad (6)$$

The two effective efficiency values $\eta_{e1}$ and $\eta_{e2}$ are considered compatible if $E_n \leq 1$.

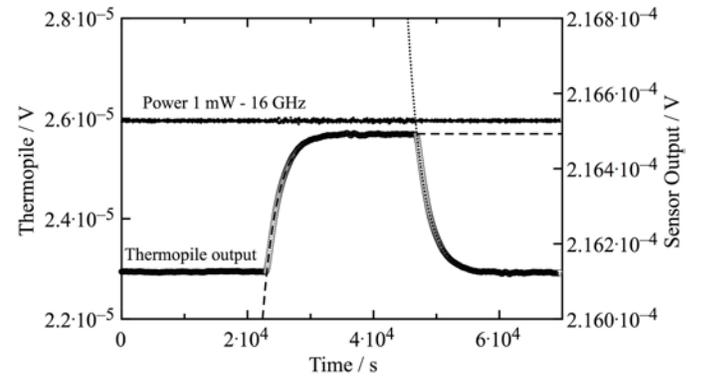

Fig. 3.   Left axis reports a typical microcalorimeter thermopile output to a power substitution of 1 mW at 16 GHz on a thermoelectric sensor. On the right axis, the voltage output of the power sensor is shown.

TABLE I

CALIBRATION POINTS OF THERMOELECTRIC POWER STANDARD. CASE (1): POWER SUBSTITUTION ON RF LINE. CASE (2): POWER SUBSTITUTION ON AUXILIARY HEATER. $E_N$ REPRESENTS THE COMPATIBILITY INDEX

| Frequency (GHz) | $\eta_e$ case (1) | $U(\eta_e)$ $k=2$ | $\eta_e$ case (2) | $U(\eta_e)$ $k=2$ | $E_n$ |
|---|---|---|---|---|---|
| 0.05 | 0.9996 | 0.0012 | 0.9993 | 0.0013 | 0.18 |
| 1 | 0.9962 | 0.0012 | 0.9957 | 0.0013 | 0.28 |
| 2 | 0.9908 | 0.0012 | 0.9913 | 0.0013 | 0.32 |
| 3 | 0.9858 | 0.0012 | 0.9858 | 0.0013 | 0.02 |
| 4 | 0.9810 | 0.0012 | 0.9814 | 0.0013 | 0.25 |
| 5 | 0.9792 | 0.0012 | 0.9795 | 0.0013 | 0.21 |
| 6 | 0.9773 | 0.0012 | 0.9767 | 0.0013 | 0.32 |
| 7 | 0.9755 | 0.0012 | 0.9754 | 0.0013 | 0.06 |
| 8 | 0.9746 | 0.0012 | 0.9741 | 0.0013 | 0.33 |
| 9 | 0.9730 | 0.0012 | 0.9731 | 0.0013 | 0.07 |
| 10 | 0.9711 | 0.0012 | 0.9703 | 0.0013 | 0.44 |
| 11 | 0.9674 | 0.0013 | 0.9680 | 0.0014 | 0.32 |
| 12 | 0.9657 | 0.0014 | 0.9659 | 0.0015 | 0.10 |
| 13 | 0.9643 | 0.0017 | 0.9637 | 0.0019 | 0.24 |
| 14 | 0.9625 | 0.0013 | 0.9620 | 0.0014 | 0.29 |
| 15 | 0.9593 | 0.0012 | 0.9592 | 0.0013 | 0.04 |
| 16 | 0.9574 | 0.0012 | 0.9573 | 0.0013 | 0.08 |
| 17 | 0.9565 | 0.0011 | 0.9569 | 0.0013 | 0.25 |
| 18 | 0.9550 | 0.0012 | 0.9549 | 0.0013 | 0.10 |

An example of uncertainty budget is given in Table 2.

From the measurement results it is obvious that the substitution paths are equivalent. Our measurements say that $H_{dc}$ in (4) is reasonably not sensitive to the change of the power substitution paths, of course in the limit of the instrumentation sensitivity. This means that the main absorber and the auxiliary heater are both well thermally coupled to the sensor thermometer, i.e. internal thermocouple schematized in Fig. 2.

Therefore, the effect of the power substitution results independent of the mode it is performed, that is, directly on the RF/MW absorber or on the dc auxiliary heater. As a consequence (if this result will be confirmed by testing other devices), by using this kind of sensors, the realization of the RF/MW primary power standard no longer depends on the assumption of unitary efficiency at dc or LF.

IV. CONCLUSION

The realization was here shown of a RF/MW broadband primary power standard by applying non conventional measurement procedure.

Basically, we used as calorimetric load a new type of thermoelectric sensor that is fitted, directly by the manufacturer, with two electrically independent absorbers, supposed to be also thermally equivalent. This new sensor architecture allowed applying the power substitution method in two different ways.

TABLE II

DETAILS OF UNCERTAINTY BUDGET AT 18 GHz FOR THERMOELECTRIC STANDARDS (QUANTITIES AND RELATED UNCERTAINTIES ARE IN VOLT, EXCLUDING ADIMENSIONAL REFLECTION COEFFICIENT).

| Influence Variable | Measured Value $y$ | Measurement Uncertainty $u(y)$ | Sensitivity coefficient $|c(y)|$ | Uncertainty Contrib. $c(y)u(y)$ |
|---|---|---|---|---|
| *case* (1) | | | | |
| $e_1$ | 2.5873E-05 | 1.05E-08 | 4.0043E+04 | 0.00042 |
| $e_2$ | 2.2943E-05 | 9.74E-09 | 4.1777E+04 | 0.00041 |
| $e_{1SC}$ | 3.8717E-06 | 4.92E-09 | 1.9883E+04 | 0.00010 |
| $\Gamma_S$ | 0.0174 | 0.0080 | 0.00128 | 0.00001 |
| $U(\eta_e)$ $k=2$ | | | | 0.0012 |
| *case* (2) | | | | |
| $e_1$ | 2.5915E-05 | 1.05E-08 | 3.9951E+04 | 0.00042 |
| $e_2$ | 2.2979E-05 | 1.17E-08 | 4.1697E+04 | 0.00049 |
| $e_{1SC}$ | 3.6984E-06 | 4.92E-09 | 1.9982E+04 | 0.00010 |
| $\Gamma_S$ | 0.0174 | 0.0080 | 0.00135 | 0.00001 |
| $U(\eta_e)$ $k=2$ | | | | 0.0013 |

The sensor has been therefore calibrated twice in the frequency range 0.01 – 18 GHz. One calibration has been performed according to the method routinely used at INRIM, that is, by supplying the reference power directly on the input coaxial line. In a second step, the calibration has been repeated by feeding the auxiliary heater with the same amount of 1 kHz power.

The results of the calibrations are compatible in the limit of the measurement uncertainties, therefore they can be used to obtain a unique value of effective efficiency at each calibration frequency. This experimentally verifies the equivalence of the two procedures and the thermal equivalence of the two absorbers.

An even more important result is that, by using the secondary heater to supply the reference power to the sensor, the primary RF/MW power standard can now be realized independently of the assumption of unitary efficiency at dc or LF, which involves complex measurements to be verified.

Moreover, because thermoelectric sensors can now be calibrated through an auxiliary heater that is electrically independent of the main absorber, then they can be used to realize hollow waveguide power standards, where the coaxial technology cannot be easily used.

Furthermore, having experienced lack of bolometric devices on the market for long time, the microcalorimeter technique finds finally a suitable replacement for its fundamental component.

Our results need to be confirmed by measuring several other sensors, because other measurements in literature [14] seems to highlight a not perfect thermal coupling between the RF absorber and the auxiliary heater of a similar waveguide sensor. By measuring a device in 7mm coaxial line up to 18 GHz, we did not find such thermal imbalance. This finding will be verified with future measurements of course, but having this technology reached full maturity to be used in devices launched on the market, we are already quite confident in our conclusions.

ACKNOWLEDGEMENT

Authors thank the technicians D. Serazio and F. Francone for their contribution to the realization of a specific thermal case necessary to adapt Rohde & Schwarz thermoelectric power sensors to the INRIM coaxial microcalorimeter.

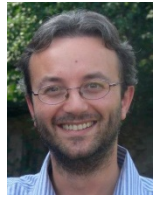

**Luca Oberto** was born in Pinerolo (Torino), Italy, on June 9, 1975. He received the M.S. degree in Physics from the University of Torino in 2003 and the Ph.D. in Metrology from the Politecnico di Torino in 2008.

From 2002 to 2003 he was with the Istituto Nazionale di Fisica Nucleare (INFN), Torino Section, working at the COMPASS experiment at CERN, Geneva, Switzerland. From 2003 he is with the Istituto Nazionale di Ricerca Metrologica (INRIM, formerly IEN "Galileo Ferraris"), Torino, Italy. His research interests are in the field of high frequency and THz metrology.

Dr. Oberto is member of the Associazione Italiana Gruppo di Misure Elettriche ed Elettroniche (GMEE). He was recipient of the 2008 Conference on Precision Electromagnetic Measurements (CPEM) Early Career Award and of the 2010 GMEE "Carlo Offelli" Ph.D. prize.

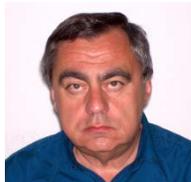

**Luciano Brunetti** was born in Asti, Italy, on September 11, 1951. He received the M.S. degree in Physics from the University of Torino, Italy, in 1977.

From 1977 to 2016 he worked at Istituto Nazionale di Ricerca Metrologica (INRIM, formerly IEN "Galileo Ferraris"), Torino, Italy. He dealth both with theoretical and experimental research in the field of high frequency primary metrology. His main task was the realization and the dissemination of the national standard of power, impedance and attenuation in the microwave range. He was also involved in the design and characterization of millimeter and microwave devices working at cryogenic temperature and he collaborated at the characterization of complex magnetic alloys at high frequency, too.